\documentstyle[prl,multicol,aps,epsfig]{revtex}
\renewcommand{\narrowtext}{\begin{multicols}{2} \global\columnwidth20.5pc}
  \renewcommand{\widetext}{\end{multicols} \global\columnwidth42.5pc}

\begin{document}
\title{Non-Fermi-Liquid in a modified single electron
transistor}
\author{Yuval Oreg$^1$ and David Goldhaber-Gordon$^2$ \date{\today}}
\address{
$^1$ Department of Condensed Matter Physics, Weizmann Institute of Science, Rehovot, 76100, ISRAEL \\
$^2$ Geballe Laboratory for Advanced Materials and Department of
 Physics, Stanford University, Stanford, CA 94305, USA}
\maketitle

\begin{abstract}
 At low temperatures, a system built from a small droplet of
electrons and a larger, but still finite, droplet may display
non-Fermi-liquid behavior. Stabilization of a multi-channel Kondo
fixed point requires fine control of the electrochemical potential
in each droplet. The desired fine control can be achieved by
adjusting voltages on nearby gate electrodes. We study the
conditions for obtaining this type of non-Fermi-liquid behavior
and discuss the experimentally-observable consequences.
\end{abstract}

\narrowtext

\begin{figure}[h]
  \vglue -0.3cm \epsfxsize=0.9\hsize
\centerline{\epsffile{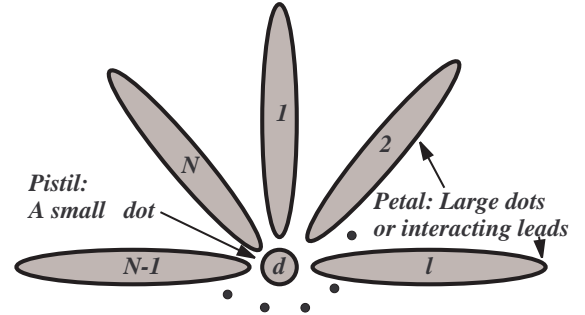}} \caption{\label{fg:flower} A
few large, but finite, conducting droplets (large dots) tunnel
coupled to a small one-level droplet (dot $d$).}
\end{figure}

Despite the presence of strong electron-electron interactions,
most ``metallic'' conductors are well-described by Landau`s
phenomenological Fermi-liquid (FL) theory~\cite{Pines66,AGD63}.
Among the few exceptions are various types of
superconductors~\cite{AGD63}, Laughlin liquids in two dimensional
electron systems at strong magnetic field~\cite{Prange87}, and
Luttinger liquids in one dimension~\cite{Fradkin91}. In this
Letter we argue that a simple configuration of two or more large
electron droplets (see Fig.~\ref{fg:flower}) attached to a small
electron droplet can exhibit multi-channel Kondo (MCK)
correlations~\cite{Affleck93}, retaining non-Fermi-liquid (NFL)
behavior at low temperature.

The single-channel Kondo (1CK) effect has been studied for decades
in metals with magnetic impurities~\cite{Hewson93}. The same
phenomenon was observed recently in semiconductor nanostructures
containing no magnetic impurities: instead, in each study an
electron droplet with a degenerate ground state plays the role of
a magnetic impurity, and nearby electron reservoirs play the role
of the surrounding normal
metal~\cite{Goldhaber98,Cronenwett98,Goldhaber98a,Wiel00,Nygard00,Liang02}.

\begin{figure}[h]
\vglue -0.2cm \hglue -0.25cm \epsfxsize=0.9\hsize
\centerline{\epsffile{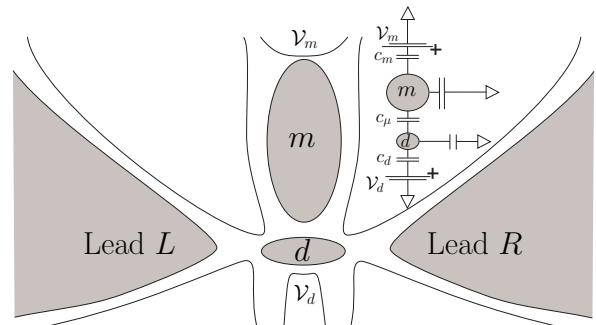}}
\caption{\label{fg:2-Coulomb-islands} A proposed realization of
the two-channel Kondo (2CK) model. Two non-interacting leads ($L$
and $R$) and a large dot~$m$ are attached to a single-level small
dot~$d$. If dot~$d$ is occupied by a single electron, it can flip
its spin by virtually hopping the electron onto either dot~$m$ or
the leads, and then returning an electron with opposite spin to
dot~$d$. Dot~$m$ and the leads thus serve as the two distinct
screening channels required to produce the 2CK effect. Crucially,
when $T$ is smaller than the charging energy of dot~$m$, Coulomb
blockade blocks transfer of electrons between the leads and
dot~$m$. Fine tuning of ${\cal V}_m$ (and/or ${\cal V}_d$) can
equalize the coupling to the two channels, stabilizing the 2CK
fixed point.}
\end{figure}

 In contrast, due to an intrinsic channel anisotropy the MCK
effect may not be observable in metals with magnetic
impurities~\cite{Nozieres80a}, making it an even more intriguing
phenomenon to produce in artificial
nanostructures~\cite{footnote02a,Ralph2levels,Matveev95,Rosch01}.
 We will show below that in a certain configuration of electron
droplets (Fig.~\ref{fg:2-Coulomb-islands}), tuning the voltage on
just one gate electrode can stabilize the two-channel Kondo (2CK)
fixed point at low temperature. At this fixed point the
conductance through the small droplet should have an anomalous
power-law dependence on temperature, a manifestation of NFL
behavior. The simplicity of the structure and the ability to tune
system parameters hold out hope for detailed study of the NFL
realm, including non-thermodynamic quantities such as transmission
phase~\cite{Gerland00,Ji00} and tunneling density of states.

To see how NFL behavior can be realized in a system of electron
droplets, we first discuss a model in which a few large conducting
droplets are attached to a single small droplet (see
Fig.~\ref{fg:flower}).
A small central ``pistil'' droplet (denoted by $d$) has a single
level of energy $\varepsilon_{ds}$, which can be empty, or
occupied by electrons of either or both spin directions
$s=\uparrow, \downarrow$. Henceforth we refer to this droplet as
{\em small dot d}. In the large ``petal'' droplets we neglect the
discreteness of single-particle energy levels, while retaining a
finite Coulomb energy. Thus, they  behave as ``interacting
leads''; we refer to them as {\em large dots}.

To describe this system we use the model Hamiltonian:
\begin{eqnarray}
 H&=&\sum_{k l s} \varepsilon^{\vphantom 0}_{l k s} l^\dagger_{k s}
l^{\vphantom \dagger}_{k s} + \sum_s \varepsilon_d^0 d^\dagger_s
d^{\vphantom \dagger}_s+ U n^{\vphantom \dagger}_{d\uparrow}
n^{\vphantom \dagger}_{d \downarrow} \nonumber\\ &+& \sum_l
u^{\vphantom \dagger}_l (n^{\vphantom \dagger}_l - {\cal
N}^{\vphantom \dagger}_l)^2+ \sum_{k s} \label{eq:model} \left(
V^*_{l k} l^\dagger_{ ks } d^{\vphantom \dagger}_s+
V^{\vphantom{*}}_{l k} d^\dagger_s l^{\vphantom \dagger}_{ks}
\right).
\end{eqnarray}
Here $l^{\vphantom \dagger}_{ks}$ is the annihilation operator of
an electron at state $k$, spin $s$ and energy
$\varepsilon^{\vphantom 0}_{lks}$ at large dot~$l=1,\dots,N$,
$d^{\vphantom \dagger}_s$ is the annihilation operator of an
electron with spin $s$ at small dot~$d$, $n^{\vphantom
\dagger}_l=\sum_{ks}l^\dagger_{ks} l^{\vphantom \dagger}_{ks}$ is
the number operator at dot~$l$,  and $n_{d \uparrow (\downarrow)}
= d^\dagger_{\uparrow(\downarrow)} d^{\vphantom
\dagger}_{\uparrow(\downarrow)}$. The parameter ${\cal
N}^{\vphantom \dagger}_l$ sets the equilibrium occupancy of dot
$l$, and $\varepsilon_d^0$ is the bare energy of level $d$. Each
of these can be independently tuned by adjusting potentials on
nearby gates.

Glazman and Raikh showed that a local site coupled to two
noninteracting leads maps to a 1CK problem, in which only an even
linear combination of the electron creation and annihilation
operators in the two leads couples to the local
spin~\cite{Glazman88}.  This result  generalizes straightforwardly
to arbitrary number of leads in the absence of Coulomb charging
($u_l=0$ for all $l$). The well-known 1CK system displays an
interesting many-body resonance, however, at very low temperatures
we can describe it simply as a FL superimposed with a
resonance~\cite{Nozieres74}.

A nonzero $u^{\vphantom \dagger}_l$ induces interactions among
some of the transformed lead operators, substantially changing the
low-temperature behavior of the system. In terms of the original
large dots the physical role of the $u^{\vphantom \dagger}_l$
terms is clear: at low temperatures they forbid processes in which
charge is ultimately transferred from one large dot to another.
Spin flip events --- {\it e.g.} when an electron hops onto the
small dot and then an electron with an opposite spin hops off the
small dot to the same large dot --- remain possible and may lead
to a MCK fixed point.

To see this, we study the case $\varepsilon^0_d
(\varepsilon_d^0+U)<0$,  and integrate out in the renormalization
group (RG) sense all energies up to a cutoff $\tilde D \sim a
|\varepsilon^{\vphantom 0}_d | \lesssim | \varepsilon^{\vphantom
0}_d |,\;|\varepsilon^{\vphantom 0}_d + U|$, with
$\varepsilon^{\vphantom 0}_d = \varepsilon_d^0+ \Gamma/\pi \log (a
D/|\varepsilon^{\vphantom 0}_d| )$. Here $\Gamma$ is the width of
level~$d$ in the absence of interactions, $D$ is a cutoff of order
the Fermi energy, $a$ is a number of order 1, and we have assumed
$\varepsilon^{\vphantom 0}_{d\uparrow}=\varepsilon^{\vphantom
0}_{d\downarrow}\equiv \varepsilon_d$~\cite{Haldane78}. Now we can
safely perform the Schrieffer-Wolff
transformation~\cite{Schriefer66} and find:
\begin{eqnarray}
 H&=&\sum_{k l s} \varepsilon^{\vphantom 0}_{l k s} l^\dagger_{k s}
l^{\hphantom{\dagger}}_{k s} + \sum_l u^{\vphantom \dagger}_l (n^{\vphantom \dagger}_l - {\cal N}^{\vphantom \dagger}_l)^2  \nonumber \\
&+& \sum_{lm, k q} J^{kq}_{lm}\left[ S_{\vphantom k}^+
s^{-kq}_{lm}+S_{\vphantom k}^- s^{+kq}_{lm}+2 S_{\vphantom k}^z
s^{zkq}_{lm} \right],\label{eq:SW}
\end{eqnarray}
where
 $ S^{\pm}_{\vphantom q}=d^\dagger_{\uparrow(\downarrow)} d^{\vphantom \dagger}_{\downarrow (\uparrow)}$,
 $S_{\vphantom k}^z=\frac{1}{2} \left(d^\dagger_{\uparrow}
d^{\vphantom \dagger}_\uparrow-d^\dagger_{\uparrow} d^{\vphantom
\dagger}_\uparrow \right)$, $s^{\pm kq}_{lm}=l^\dagger_{k\uparrow
(\downarrow)} m^{\vphantom \dagger}_{q \downarrow(\uparrow)}$, $
s^{zkq}_{lm}=\frac{1}{2} \left(l^\dagger_{k\uparrow} m^{\vphantom
\dagger}_{q \uparrow}-l^\dagger_{k \downarrow} m^{\vphantom
\dagger}_{q \downarrow} \right)$ and
\begin{equation}
\label{eq:Jlm} J_{lm}^{kq}=V^{\vphantom m
}_{lk}V_{mq}^*\left[\frac{1}{E_{\rm elec}^{qm}-E^{\vphantom
m}_{\rm init}}+\frac{1}{E^{kl}_{\rm hole}-E^{\vphantom m}_{\rm
init}} \right].
\end{equation}
Here $E_{\rm init}$ is the energy of the initial state with one
electron in the small dot~$d$ and occupancy $n_{l}\; (n_{m})$ in
large dot~$l\;(m)$; $E^{qm}_{\rm elec}$ denotes the energy of an
intermediate state of a process where first an electron hops from
state $q$ of dot~$m$ onto dot~$d$ and then the same electron or an
electron with opposite spin hops off dot~$d$ onto state $k$ of
dot~$l$; $E^{kl}_{\rm hole}$ is the energy of the intermediate
state of a process where the temporal order of the hopping events
is interchanged.

Let us assume that $\varepsilon^{\vphantom 0 }_{l(m)qs} \approx
\varepsilon^{\vphantom 0}_{l(m)F}$, where $\varepsilon^{\vphantom
0 }_{l(m)F}$ is the energy of the last empty (occupied) state in
dot~$l\;(m)$, and that $V_{mq}=V_{lk}\equiv V$. Then using
Eq.~(\ref{eq:model}) to express the energies in Eq.~(\ref{eq:Jlm})
we find:
\begin{equation}
\label{eq:J1} J_{lm}^{kq}=J_{lm}=|V|^2 \frac{U+u^{-}_m+u^{+}
_l}{\left[U+\varepsilon_d+u^{-}
_m\right]\left[u^{+}_l-\varepsilon_d\right]},
\end{equation}
where $u^{\rm \pm}_{p}= u_{p}\left[1\pm 2\left(n_{p}-{\cal
N}_{p}\right)\right]\pm \varepsilon^{\vphantom 0}_{pF}$, and for
$-1/2 < {\cal N}_{p}-n_{p}+(\mu-\varepsilon^{\vphantom 0}_{pF})/(2
u_{p}) < 1/2$ there are $n_{p}$ electrons in dot~$p$, where
$p=m,l$ and $\mu$ is the electrochemical potential of a reference
reservoir. Particle hole symmetry may be absent in the large dots,
so in general $J_{lm} \neq J_{ml}$.

 The off-diagonal terms describe transfer of an electron from
 dot~$m$ to dot~$l$; at low temperature these
processes are exponentially suppressed as $J^{\vphantom 0 }_{lm}
=J_{lm}^0 e^{u^{\vphantom 0 }_{lm}/(4T)}$ with $ u^{\vphantom 0
}_{lm}= \left(u^+_l+ u_m^-\right) \left(1- \delta_{lm} \right)$.
Notice that $u^{\vphantom 0 }_{ll}=0$, since the charge on dot~$l$
is not changed when an electron hops from dot~$l$ onto dot~$d$ and
then back to the same dot~$l$. At $T < \min(u^{\vphantom 0
}_{lm})$ only the diagonal terms of $\tilde J_{lm}$ do not flow to
zero. Assuming that we are not at a degeneracy point where
$u^{\vphantom 0 }_{lm}=0$, an easy condition to avoid, the RG
equations are identical to the MCK RG equations~\cite{Nozieres80}:
\begin{equation}
\label{eq:2CK} \frac{d \tilde J^{\vphantom 2}_{ll}}{d \log(\tilde
D/T)}= \tilde J_{ll}^2 -\tilde J^{\vphantom 2}_{ll} \sum_{p}
\tilde J_{pp}^2 .
\end{equation}

As in the case of classic MCK, our NFL fixed point is unstable to the
introduction of channel anisotropy. If one of the coupling constants
is larger than the others, the corresponding channel alone screens the
local spin and forms a Kondo resonance while the other channels are
decoupled from the local spin. In our model we can tune ${\cal N}_l$
to achieve $J_{ll}=J$ for all $l$. For $N$ large dots this requires
tuning of $N-1$ gate potentials.

These gate potentials capacitively control the energy of the last
occupied level in each large dot, so excitations in each large dot
will be around a different Fermi energy. This does not modify the
RG equations, but will affect certain physical properties such as
the small dot density of states at finite energies. A similar
situation occurs in the discussion of 2CK in a dot out of
equilibrium~\cite{Rosch01,footnote02a}.

For large $N$ it would be difficult to realize MCK
experimentally, requiring tuning at least $N-1$ gate potentials to
obtain the desired fixed point. We therefore propose a specific
realization of the 2CK model (Fig.~\ref{fg:2-Coulomb-islands}). In
this ``three leg'' structure two noninteracting ``free'' leads $L$
and $R$ --- effectively large dots with $u_{L,R}=0$ --- are
connected to small dot~$d$ which is in turn connected to large
dot~$m$. Dot~$m$ is not directly connected to the leads. This
structure allows conductance measurements between leads $L$ and
$R$ through dot $d$, and proper tuning of parameters
should cause 2CK effect to proclaim itself in this conductance.

\begin{figure}[h]
 \vglue 0cm \hglue -0.25cm \epsfxsize=0.9\hsize
\centerline{\epsffile{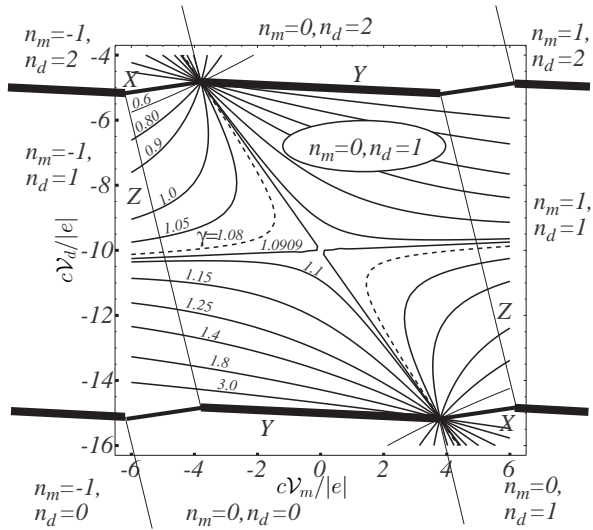}} \vspace{-.2cm}
\caption{\label{fg:hexagons} The number of electrons on dots $m$
and $d$ as a function of the gate potentials ${\cal V}_{m}$ and
${\cal V}_{d}$ (see Fig.~2 and [26]). Transitions between
configurations with identical $n_m$ but different $n_d$ (marked by
the letter $Y$) occur by transfer of one electron from dot~$d$ to
the leads, and those in which $n_m$ increases by one and $n_d$
decreases by one or vice-versa ($X$) occur by transfer of an
electron between dots $m$ and $d$. By contrast, the thin lines
($Z$) across which only $n_m$ changes are reminders that there is
no direct coupling between dot~$m$ and the leads, making direct
transitions across these lines difficult.
 The curves superimposed on the central hexagon (``2CK-lines'')
map where in the ${\cal V}_m, {\cal V}_d$ plane the two-channel
Kondo (2CK) effect is realized. Each value of the coupling ratio
$\gamma\equiv\Gamma_m/\Gamma_{l}$ gives rise to a pair of disjoint
curves, as illustrated for $\gamma=1.08$ (dashed) [26]. These two
curves divide the hexagon into three regions with distinct
low-temperature fixed points. On the curves the 2CK effect is
realized and the deviation of the interlead differential
conductance from its $T \rightarrow 0, {\cal V}_{LR} \rightarrow
0$ limit $G(0,0)$ is $\propto \sqrt{\max\left(T,{\cal
V}_{LR}\right)}\!$~, [Eq.~(\ref{eq:2CKwins})]. In regions
adjoining the~$X$ boundaries dot~$m$ ``wins'', forming a 1CK
resonance with dot~$d$, and driving $G(T,{\cal V}_{LR})$ close to
zero [Eq.~(\ref{eq:Largedotwins})]. In regions adjoining the $Y$
boundaries, leads $L$ and $R$ ``win'' giving rise to Fermi-liquid
behavior $G(0,0)-G(T,{\cal V}_{LR}) \propto
\left[\max\left(T,{\cal V}_{LR}\right)\right]^2$,
[Eq.~(\ref{eq:Leadswin})].}
\end{figure}
\vspace{-0.3cm}
Let the potential ${\cal V}_{d}$ on the gate near dot~$d$ be such
that dot~$d$ is singly-occupied in the absence of coupling to dot
$m$. Then if $\Gamma_m$, the bare coupling between
dots~$d$~and~$m$, is roughly equal to $\Gamma_l$, the coupling
between dot~$d$ and the leads, fine-tuning the gate potential
${\cal V}_{m}$ can stabilize the 2CK effect.
 To demonstrate this, we analyze the electrostatics of the
circuit shown in the inset of Fig.~\ref{fg:2-Coulomb-islands}.

For fixed number of electrons $n_{m}\; (n_{d}) $ on dots $m\;
(d)$, the electrostatic energy of the system is:
\begin{eqnarray}
 E_{\rm ele}^{n_m,n_d}&\equiv& E_{\rm ele}^{n_m,n_d}
 \left({\cal V}_m,{\cal V}_d\right)  \nonumber \\
   &=& U \left(n_{d}- {\cal N}_d \right)^2
  +  u_m\left(n_{d}+ \alpha n_{m}-{\cal N}\right)^2, \label{eq:electrostatics}
\end{eqnarray}
where $U \equiv e^2/(2 \tilde C_d) \gg u_m \equiv e^2/(2 \tilde
C_m-c_\mu^2/\tilde C_d)$, $ |e| {\cal N}_{d} \equiv  c_d {\cal
V}_{d}$, $|e| {\cal N} \equiv c_m {\cal V}_{m}+c_d c_\mu/\tilde
C_m {\cal V}_{d}$ and $\alpha \equiv c_d c_\mu /\tilde C_d$. Here
$\tilde C_{m (d)}$ is the {\em total} capacitance of dot~$m (d)$.
See Fig.~\ref{fg:2-Coulomb-islands} for definitions of the other
capacitances.

Since dot~$m$ is large we assume that $\tilde C_m$ is much larger
than all other capacitances. The number of electrons on each dot
can be estimated by minimizing $E_{\rm ele}$ with respect to
$n_{m}$ and $n_{d}$~\cite{footnote02c}. In Fig.~\ref{fg:hexagons}
we plot the number of electrons on each dot as a function of
${\cal V}_{m}$ and ${\cal V}_{d}$~\cite{footnote02b}.

To write down the full Hamiltonian $\tilde H$ of the model system
we note that there is no Coulomb blockade for transfer of
electrons between $L$ and $R$ leads, so we can define $ {l}_{k
s}=\cos \theta \;L_{ks} + \sin \theta \;R_{ks},\;\; o_{k s}=\cos
\theta \;R_{ks} - \sin \theta \;L_{ks},\;\; \tan \theta =
V_R/V_L,\;\; V_{l} = \sqrt{|V_L|^2+|V_R|^2}$. Without loss of
generality we take the coupling constants $V_i$, $i=L,R,m$, to be
real. With these definitions:
\begin{eqnarray}
\tilde H &=& \sum_{i=o,m,l;ks} \epsilon^{\vphantom \dagger}_{iks}
i^\dagger_{ks} i^{\vphantom \dagger}_{ks} +\sum_s
\epsilon^{\vphantom \dagger}_{d s} d^\dagger_s d^{\vphantom \dagger}_s + E_{\rm ele}^{n_m,n_d}({\cal V}_m,{\cal V}_d) \nonumber \\
&&+V_m \sum_{ks} m^\dagger_{ks} d^{\vphantom
\dagger}_s+V_{l}\sum_{ks} {l}^\dagger_{ks} d^{\vphantom \dagger}_s
+ {\rm h.c.},\label{eq:H2cka}
\end{eqnarray}
 which is similar to  Eq.~(\ref{eq:model}). To obtain a 2CK
fixed point we tune ${\cal V}_{m}$ and ${\cal V}_{d}$ to make
$\tilde J_{mm}$, the coupling of dot~$d$ to dot~$m$,  equal to
$\tilde J_{ll}$, the coupling of dot~$d$ to the leads:
\begin{eqnarray}
 \tilde J_{mm}({\cal V}_m,{\cal V}_d)&\equiv& \Gamma_m \left[ \frac{1}{E_{\rm ele}^{1,0}- E_{\rm ele}^{0, 1} }+ \frac{1}{E_{\rm ele}^{-1,
2}- E_{\rm ele}^{0, 1}}\right] =  \nonumber   \\
\tilde J_{ll}({\cal V}_m,{\cal V}_d)&\equiv&\Gamma_{l} \left[
\frac{1}{E_{\rm ele}^{1,0}- E_{\rm ele}^{0, 2} }+ \frac{1}{E_{\rm
ele}^{0, 0}- E_{\rm ele}^{0, 1}}\right]. \label{eq:condition}
\end{eqnarray}
Here $\Gamma_{m(l)} = |V_{m(l)}|^2 \nu_{m(l)}$, where
 $\nu_{m(l)}$ is the density of states in dot~$m$ (the leads).
Eq.~(\ref{eq:condition}) together with the ratio
$\gamma\equiv\Gamma_{m}/\Gamma_{l}$ defines a curve in the ${\cal
V}_{m},{\cal V}_{d}$ plane. In Fig.~\ref{fg:hexagons} we show
several of these ``2CK-lines'' for different values of $\gamma$.
On these lines 2CK physics should be realized at low $T$.

The 2CK fixed point can be reached experimentally by fixing ${\cal
V}_d$ to give an odd number of electrons in the small dot, and
fine tuning ${\cal V}_m$ so that $\tilde J_{mm}({\cal V}_m,{\cal
V}_d)= \tilde J_{ll}({\cal V}_m,{\cal V}_d)$. In
Fig.~\ref{fg:hexagons} this corresponds to tuning ${\cal V}_m$
until we hit the 2CK line for our given value of $\gamma$.

The current $I$ between the left and right leads can be measured
as a function of temperature $T$, and as a function of ${\cal
V}^{\vphantom 0}_{LR}$, the bias applied between leads $L$ and $R$
(see Fig.~\ref{fg:2-Coulomb-islands}). Drawing on the extensive
literature of 2CK physics~\cite{Cox98} we can predict the
qualitative behavior of the $I$-${\cal V}^{\vphantom 0}_{LR}$
curve through the dot, for different values of the gate voltages
${\cal V}_m$, ${\cal V}_d$ that scan the the hexagon of
Fig.~\ref{fg:hexagons}.

The 2CK-lines divide the hexagon into three different parts.
  On the 2CK-lines in the unitary limit
--- $T$,${\cal V}_{RL} \ll$ Kondo temperature $T_K \cong \sqrt{U
\Gamma_m} e^{-1/\tilde J_{mm}({\cal V}_m,{\cal V}_d)}$
--- the differential conductance $G(T,{\cal V}^{\vphantom 0}_{LR}) \equiv dI/d{\cal V}^{\vphantom 0}_{LR}$ should
approaches its limiting value $G(0,0)$ as
\begin{mathletters}
\begin{eqnarray}
\label{eq:2CKwins}
 2CK:\;\;  G(0,0)-G(T,{\cal V}^{\vphantom 0}_{LR})\propto
\sqrt{\max({\cal V}^{\vphantom 0}_{LR},T)}.
\end{eqnarray}
In the symmetric case $V_L=V_R$, we get $G(0,0) = G_K \equiv
e^2/(2\pi\hbar)$, half the maximal value of $G(0,0)$ in the 1CK
effect~\cite{Wiel00,Liang02}.

In regions of the hexagon adjoining boundaries labelled $X$ in
Fig.~\ref{fg:hexagons}, ({\em i.e.} for $\tilde J_{mm} > \tilde
J_{ll}$), as temperature decreases the electrons in dot~$m$ screen
the spin of dot~$d$, while the leads are decoupled. In the RG
sense $\tilde J_{ll}$ flows to zero, so that  dot~$m$ ``wins''
over the leads and forms a 1CK state with dot~$d$. In that case $d
I /d {\cal V}^{\vphantom 1}_{LR}$ is small and given by
\begin{equation}
\label{eq:Largedotwins} \text{ Large dot wins:}\;\;G(T,{\cal
V}^{\vphantom 0}_{LR}) \propto \left[\max\left({\cal V}^{\vphantom
0}_{LR},T\right)\right]^2\!\!.
\end{equation}

 In contrast, in regions of the hexagon adjoining boundaries
labelled $Y$ in Fig.~\ref{fg:hexagons} ({\em i.e.} for $\tilde
J_{mm} < \tilde J_{ll}$), dot~$m$ decouples from dot~$d$ at low
temperature, leaving the leads to form a 1CK resonance with dot
$d$ and
\begin{equation}
\label{eq:Leadswin} \!\!\text{Leads win:}\;\;G(0,0) - G(T,{\cal
  V}^{\vphantom 0}_{LR})\propto \left[\max \left({\cal
  V}^{\vphantom 0}_{LR},T\right)\right]^2\!\!\!,
\end{equation}
\end{mathletters}
where $G(0,0)=2G_K$ for $V_L=V_R$.

At sufficiently low temperature the finite level spacing
$\Delta_m$ in dot~$m$ will cut off the RG flow of the coupling
constants~\cite{Thimm99}. We cannot make $\Delta_m$ infinitesimal
as we must retain a finite Coulomb blockade energy $u_m > kT$.
However, the ratio between charging energy and level spacing can
be made large, allowing 2CK behavior to be observed over a
significant temperature range before the system finally flows to
the 1CK fixed point.

In conclusion, if a small dot is coupled to multiple electron
reservoirs (large dots) Coulomb blockade suppresses
inter-reservoir charge transfer at low temperatures. In the
simplest version of this system, electrostatic gates provide the
tunability needed to stabilize a 2CK fixed point, resulting in
observable NFL behavior. Softer versions of suppressing
inter-reservoir tunneling could also work in place of Coulomb
blockade. For example, the reservoirs could be conductors with
large impedance~\cite{Devoret90}, one-dimensional Luttinger
liquids~\cite{Kim01} or conductors with strongly-interacting
charge carriers. Finally, while the channel asymmetry parameter is
relevant in the RG sense, for realistically well-matched channel
couplings we expect that the system will remain near the 2CK fixed
point, and will show NFL behavior, over a wide range of
temperatures.

 It is our pleasure to acknowledge useful discussions with Natan
Andrei, Michel Devoret, Gabi Kotliar, Leo Kouwenhoven, Yigal Meir,
Arcadi Shehter, Alex Silva Ned Wingreen, and especially Avi
Schiller. This work was supported by DIP grant c-7.1, by ISF grant
160/01-1, and by Stanford~University.

\widetext

\begin{thebibliography}{10}
\bibitem{footnote02a}
We emphasize that our proposal differs from other suggested
experimental
  realizations of the 2CK effect~\protect{\cite{Ralph2levels,Matveev95}}, in
  that our local degeneracy comes from spin rather than charge or some other
  two-level system. Our mapping differs from that of~\protect{\cite{Matveev95}}
  in several essential ways: we consider a system with fixed electron number,
 not at a charge degeneracy point; our Kondo temperature is set by the charging energy of a small dot;
 and we do not require single-mode coupling between the dot and leads.
 Our 2CK effect is realized at equilibrium, avoiding decoherence and
 suppression of 2CK by voltage-driven current (\cite{Rosch01} and
 references therein.)
\bibitem{footnote02c}
In principle we should also include the kinetic terms to determine
the number
  of particles on the dot, however these may be absorbed in the definitions of
  ${\cal N}_d, {\cal N}_a$.
\bibitem{footnote02b}
In Fig.~\protect{\ref{fg:hexagons}} we take $c_d=c;\; \tilde
C_d=4c;\; c_\mu=c;
  \; c_m=c; \; \tilde C_d=22 c$, and measure voltages in units of $e/c$. The
  exact values of parameters have no major significance so long as
  $\tilde C_m$ is much larger than all other capacitances.
\end{thebibliography}
\end{document}